\documentclass[twocolumn]{aastex631}

\newcommand{\revI}[1]{\textrm{#1}}
\newcommand{\revII}[1]{\textrm{#1}}
\newcommand{\revIII}[1]{\textrm{#1}}
\usepackage{bm}
\usepackage{csquotes}
\usepackage{subfiles}
\usepackage{amsmath}
\usepackage{comment}

\makeatletter
\def\footnoterule{\kern-3\p@
  \hrule \@width 2in \kern 2.6\p@} % the \hrule is .4pt high
\makeatother

\makeatletter
\patchcmd{\ltx@foottext}{%
  .5\textwidth\advance\hsize-18pt}{%
  \linewidth\advance\hsize-1.8em%
}{}{}
\makeatother

\defcitealias{RJW_1982}{RJW}

\begin{document}

\title{The Effects of Magneto-Convection on Short Period Cataclysmic Variables}

\author[0000-0002-4524-9497]{Conor M. Larsen}
\affiliation{Department of Physics and Astronomy,
University of Delaware, Newark, DE, 19716, USA}

\author[0000-0002-0506-5124]{James MacDonald}
\affiliation{Department of Physics and Astronomy,
University of Delaware, Newark, DE, 19716, USA}

\begin{abstract}

Many of the current problems related to the evolution of cataclysmic variables revolve around the magnetic nature of the main sequence secondary. It is known that magnetic fields alter the structure of low mass stars. In particular, they inhibit convection, leading to inflated radii. Here we present a simple model to demonstrate the impact of magneto-convection on the evolution of short period cataclysmic variables. We find that the inclusion of magneto-convection leads to larger secondaries, longer orbital periods and smaller mass-loss rates. When including magnetic effects, the minimum orbital period is increased by 14 minutes, indicating that this could help alleviate the period minimum problem in cataclysmic variable evolution. We also examine the effect of the white dwarf mass on the minimum period. While increasing the white dwarf mass does increase the minimum period, it is not substantial. Therefore it is unlikely that the period minimum problem can be solved with a larger white dwarf mass or with mass growth of white dwarf.

\end{abstract}

\keywords{Cataclysmic variable stars (203), Close binary stars (254), Magnetic stars (995),  Stellar evolution (1599)}

\section{Introduction} \label{sec:intro}
Cataclysmic variables (CVs) are semi-detached binary systems consisting of a main sequence (MS) secondary transferring material onto a white dwarf (WD) primary \citep{Warner_text}. The evolution of CVs is driven by angular momentum loss (AML). CVs with orbital periods greater than $\sim3$ hours experience AML through gravitational radiation and magnetic braking caused by the stellar wind of the secondary. Observationally, there is a dearth of CVs found with orbital periods between $\sim2-3$ hours known as the period gap \citep[see][for a recent examination of the period gap]{Period_gap}. The classical explanation of CV evolution and the period gap is the disrupted magnetic braking model \citep{RVJ}. At a period of $\sim3$ hours, the MS secondary becomes fully convective. At this point magnetic braking either ceases completely or at least becomes diminished in strength, therefore reducing the AML rate. Mass transfer initially drives the secondary out of thermal equilibrium, \revII{causing the star to be larger in size}. When the AML rate lessens, there is a sharp drop in the mass transfer rate, allowing the secondary to regain thermal equilibrium. The relaxation of the secondary pulls the star back inside its Roche lobe, cutting off mass transfer completely. AML continues through gravitational radiation. At a period of $\sim2$ hours, the Roche lobe of the secondary will have shrunk enough to resume mass transfer, although at a much lower rate since AML is driven solely by gravitational radiation. \revII{Below the gap}, systems continue to evolve to shorter periods until reaching the period minimum. \revI{At the period minimum, the secondary has been driven out of thermal equilibrium, such that the Kelvin-Helmholtz (or thermal) timescale is longer than the mass-loss timescale. The radius of the secondary can no longer adjust rapidly enough as the mass decreases, leading to an increase in the orbital period \citep{timescale_P_min}. At low masses, the secondary also becomes electron degenerate, where the response to mass loss is expansion rather than contraction. CVs that have evolved past the period minimum are known as period bouncers.}

While this picture of CV evolution is generally supported by observations, there are some major problems. One, there is much evidence that fully convective stars can sustain magnetic fields. This evidence comes from measurements of the ratio of X-ray luminosity to bolometric luminosity ($L_{X}/L_{\text{bol}}$), a known proxy for magnetic dynamo behavior \citep{full_conv_B_field_indirect} and now, even direct detections of magnetic fields in fully convective stars \citep[e.g.][]{fully_conv_B_field}. In response, \cite{Garraffo} have attempted to explain the period gap with an increase in magnetic complexity at the fully convective transition. The increase in magnetic field complexity reduces the number of open field lines which decreases the AML rate due to magnetic braking. However, a recent paper \citep{garaffo_refute} has \revI{challenged} the particular magnetic braking prescription developed by \cite{Garraffo}. Tension still remains over the appropriate magnetic braking prescription for CVs. Regardless, the mounting evidence of magnetic fully convective stars indicates that CVs \revII{below the period gap} may still retain their magnetic fields albeit at a reduced strength of magnetic braking. \revI{The decrease in AML rate across the period gap is corroborated observationally through measurements showing that the mean mass accretion rate for systems below the period gap is smaller than above the period gap \citep{ave_WD_mass}.}

Another problem relevant to this study is the orbital period minimum problem. Theoretical calculations of the period minimum have resulted in values of $\sim70$ minutes \citep[e.g.][]{RJW_1982, Kolb_P_min}. This is significantly shorter than the observed value of $80-86$ minutes, identified from SDSS data \revII{\citep{P_min_SDSS, McAllister}}. Increasing the AML rate below the period gap has been shown to increase the theoretically predicted period minimum. \cite{Knigge} have shown that multiplying the AML rate due to gravitational radiation by $2.47\pm0.22$ brings the theoretical period minimum into agreement with the observed value. \revIII{Additionally, it has been shown that the inclusion of magnetic braking below the period gap can help explain the period minimum problem \citep{Knigge, Sarkar_Tout_2022, MB_under_the_gap}.} AML due to the mass transfer process, known as consequential angular momentum loss (CAML), has also been introduced as an additional AML mechanism that can help explain the period minimum discrepancy \citep{CAML}. The physical mechanism behind CAML is still not understood, however frictional angular momentum loss following novae outbursts \citep{Jim_FAML, FAML} \revII{or binary-driven mass loss during novae outbursts \citep{Tang_2024} are possibilities.}

Here we present an alternative explanation for the period minimum discrepancy: the magnetic inhibition of convection. It has long been known that magnetic fields can alter the internal structure of low mass stars \revII{through the inhibition of convection.} Specifically, models including these magnetic effects are inflated in radius and can retain radiative cores at lower masses than standard, non-magnetic models predict \citep{MM01, MM24}. Both of these effects are relevant for the MS secondaries present in CVs. In this paper we present a simple model to demonstrate the impacts of magneto-convection on CVs below the period gap. This paper includes a \revII{description of how magneto-convection is implemented in 1D stellar evolution codes (Section \ref{sec:MC})}, a description of the model (Section \ref{sec:Model Description}), the results of the simulations (Section \ref{sec:Results}), a discussion of the results (Section \ref{sec:discussion}) and a conclusion (Section \ref{sec:conclusion}).

\section{Magneto-Convection} \label{sec:MC}

\revII{In this section we describe the implementation of magneto-convection in 1D stellar evolution codes.} When a medium is permeated by a \revI{uniform} magnetic field, magneto-convection occurs when $\nabla > \nabla_{\text{ad}} + \delta$, where $\nabla = d \ln T / d \ln P$, $\nabla_{\text{ad}} = \left(d \ln T / d \ln P\right)_{\text{ad}}$ and $\delta$ is the magnetic inhibition parameter:

\begin{equation}
    \delta = \frac{B_{v}^{2}}{B_{v}^{2} + 4 \pi \gamma p_{\text{gas}}},
\end{equation}
where $B_{v}$ is the vertical component of the magnetic field, $\gamma$ is the local ratio of specific heats and $p_{\text{gas}}$ is the local gas pressure \citep{GT66}. Without a magnetic field, convection occurs when the Schwarzschild criterion, $\nabla > \nabla_{\text{ad}}$, is met. Since $\delta$ is a positive quantity, the presence of a magnetic field increases the threshold for convection to start; the magnetic field \enquote{inhibits} convection. \revII{The convective energy transport can then be computed with the standard mixing length theory with the replacement $\nabla_{\text{ad}}\rightarrow\nabla_{\text{ad}} + \delta$. Additionally, pressure and energy terms in the equations of stellar structure are modified to include the magnetic pressure and energy.}

Recent models \citep[e.g.][]{MM17, MM24} fix $\delta$ at a constant value in the outer layers of the star. \revII{Then, when moving deeper into the star, the vertical field strength increases to compensate the increase in the gas pressure.} This continues until the magnetic field reaches a ceiling value $B_{\text{ceil}}$, at which the magnetic field \revI{is held} constant when moving deeper into the star. Thus, a complete magneto-convective model requires the specification of two parameters: $\delta$ and $B_{\text{ceil}}$.  

\revII{In the context of low-mass stars, magneto-convection has been used to explain the observed radius inflation of M-dwarfs \citep[e.g.][]{MM01,MM17}. \cite{Ireland_MC} have incorporated magneto-convective effects for low-mass stars in the stellar evolution code \texttt{MESA} \citep{MESA_2011, MESA_2023}. Magneto-convection has also been applied to high-mass stars, for example, in studying the suppression of sub-surface convective zones \citep{MacDonald_2019, Jermyn_Cantiello_2020}.}

\section{Model Description} \label{sec:Model Description}

\subsection{Polytrope Model of the Secondary Component}\label{Sec: polytrope model}

\revII{As a first step in understanding the effects of magneto-convection on CV evolution}, we model the low mass, fully convective MS secondaries below the period gap as $n=3/2$ polytropes, in a similar manner to \cite{RJW_1982} (hereafter \citetalias{RJW_1982}). A polytrope assumes a power law relationship between pressure and density:

\begin{equation}
    P = K \rho^{\left(n+1\right)/n},
\end{equation}
where $K$ is the polytropic constant and $n$ is the polytropic index. The dimensionless variables $\theta$ and $\xi$ are defined such that:

\begin{equation}
    \rho = \rho_{c}\theta^{n},
\end{equation}

\begin{equation}
    r = \alpha \xi,
\end{equation}
where $\rho_{c}$ is the central density, $r$ is the radial distance from the stellar center and the scale length $\alpha$ is given by:

\begin{equation}
    \alpha = \left[\frac{\left(n+1\right) K \rho_{c}^{\left(1-n\right)/n}}{4\pi G}\right]^{1/2},
\end{equation}
where $G$ is the universal gravitational constant. Inserting these dimensionless variables and combining the polytrope equation of state with the mass conservation and hydrostatic equilibrium equations of stellar structure yields the Lane-Emden equation \citep[see][for details]{chandra}:

\begin{equation}
    \frac{1}{\xi^{2}} \frac{d}{d\xi} \left(\xi^{2} \frac{d\theta}{d\xi}\right) + \theta^{n} = 0.
\end{equation}
With the boundary conditions $\theta\left(0\right) = 1$ and $\theta'(0) = 0$, where a prime denotes differentiation with respect to $\xi$, the Lane-Emden equation can be numerically solved to determine $\theta$ as a function of $\xi$.

Simple, analytic expressions exist for central and global properties of polytropic stars. For an $n= 3/2$ polytrope, the central density and pressure are: 

\begin{equation}
    \rho_{c} = 1.4032 \frac{M_{2}}{R_{2}^{3}},
\end{equation}

\begin{equation}
    P_{c} = K \rho_{c}^{5/3},
\end{equation}
where $M_{2}$ and $R_{2}$ are the mass and radius of the MS secondary. Stars modelled as $n = 3/2$ polytropes also obey a mass-radius relationship:

\begin{equation}\label{Eq: mass rad relation}
    R_{2} = \frac{K}{0.42422 G M_{2}^{1/3}}.
\end{equation}
In addition to the polytrope equation of state, we also use the ideal gas law with a term correcting for electron degeneracy:

\begin{equation}\label{ideal P}
    P = \frac{k_{B} \rho T \mathcal{D}}{m_{H} \mu},
\end{equation}
where $k_{B}$ is Boltzmann's constant, $m_{H}$ is the mass of the hydrogen atom and $\mu$ is the mean molecular weight. The parameter $\mathcal{D}$ is the electron degeneracy factor, which is the ratio of total gas pressure to ideal gas pressure. For non-relativistic degeneracy, $\mathcal{D}$ is given by \citepalias[see][and references therein]{RJW_1982}:

\begin{equation}
    \mathcal{D} = 1 + \frac{\mu \Gamma}{\sqrt{8\pi} \mu_{e} \left(1 + 0.05512 \: \Gamma\right)^{1/3}},
\end{equation}
where $\mu_{e}$ is the mean molecular weight per electron and:

\begin{equation}
    \Gamma = \frac{0.1105}{\mu_{e}} \frac{\rho}{T_{6}^{3/2}},
\end{equation}
where $T_{6}$ is the temperature in units of $10^{6}$ K. In an $n = 3/2$ polytrope, $\rho / T^{3/2}$ is constant throughout the star. Thus, the electron degeneracy factor $\mathcal{D}$ is constant throughout the star. Given \revII{$P_{c}$} and \revII{$\rho_{c}$}, equation \ref{ideal P} can be numerically solved for the central temperature. Using the expressions for the central density, central pressure and the mass-radius relation, $\rho_{c}$ and $P_{c}$ can be replaced in favor of $M_{2}$ and $R_{2}$. For the sake of computation time, we numerically solved equation \ref{ideal P} for $T_{c}$ over a dense grid covering $M = 0.03 - 0.3$ $M_{\odot}$ and $R = 0.03 - 0.3$ $R_{\odot}$. We then linearly interpolated over this grid so that the central temperature of the secondary can be quickly computed given the mass and radius of the secondary.

For the luminosity generated by the \textit{p-p} chain, we use the same expression used in \citetalias{RJW_1982}. In terms of the dimensionless polytrope variables, the nuclear luminosity is given by:

\begin{equation}\label{eq: L nuc}
\begin{aligned}
    L_{\text{nuc}} & = 6.08\times10^{5} X^{2} R_{2}^{3} T_{c6}^{-2/3} \rho_{c} \int_{0}^{\xi_{1}} f_{0} \theta^{7/3} \\
    & \times \exp\left[-33.81 T_{c6}^{-1/3}\theta^{-1/3}\right] \xi^{2} d\xi \:\: \text{erg/s}, 
\end{aligned}
\end{equation}
where $X$ is the hydrogen mass fraction, $T_{c}$ is the central temperature and $\xi_{1} = R_{2} / a = 3.6537$ is the value of $\xi$ at the first zero of $\theta$. The term $f_{0}$ is the electron screening factor. For $f_{0}$ we take the weak screening factor of \cite{Salpeter_screening}:

\begin{equation}
    f_{0} = \exp\left[0.188\zeta T_{6}^{-3/2}\rho^{1/2}\right],
\end{equation}
where

\begin{equation}
    \zeta = \left[X + Y + \left(\frac{f'}{f}\right)\left(X+\frac{Y}{2}\right)\right],
\end{equation}
\begin{equation}\label{eq: d}
    \frac{f'}{f} = \frac{1}{1 + \frac{2}{3}d},\:\:\:\:\: d = 0.3\left(X + \frac{Y}{2}\right)^{2/3}\rho^{2/3}T_{6}^{-1},
\end{equation}
and $Y$ is the helium mass fraction. The term $\zeta$ generally contains sums over all nuclear species present. We only include hydrogen and helium. \citetalias{RJW_1982} take the screening correction to be a constant multiplicative factor. However, $f_{0}$ does change throughout an $n=3/2$ polytrope, albeit only slightly in the regions where nuclear burning is substantial. We chose to keep $f_{0}$ inside the integral and not assume it constant. We numerically solved the integral in equation \ref{eq: L nuc} on a dense grid grid covering $T_{c}$ from $(1-9)\times10^{6}$ K and $\rho_{c}$ from $20-1000$ g/cm\textsuperscript{3}. Nuclear reactions were cutoff (that is, the integrand in equation \ref{eq: L nuc} was set to zero) in any region where the temperature is below $T = 10^{6}$ K. We then created a linear interpolation in order to quickly compute the integral in equation \ref{eq: L nuc} for a given value of $T_{c}$ and $\rho_{c}$.

\subsection{Evolution of the Secondary and Boundary Conditions}

In order to model the secondary components of CVs, we require an equation expressing how the polytropic constant $K$ changes with time. Starting with the conservation of energy equation of stellar structure \citep[see equation 5.3 in][]{Jim_book} and applying the polytropic equation of state and internal energy per unit mass for an ideal gas, one obtains such a relationship. The change in the polytropic constant $dK$ in a timestep $dt$ is:

\begin{equation}\label{Eq: dK}
    \frac{dK}{K} = \frac{7}{3}\frac{R_{2}}{G M_{2}^{2}} \left(L_{\text{nuc}} - L_{s}\right) dt,
\end{equation}
where $L_{s}$ is the surface luminosity calculated with:

\begin{equation}\label{Eq: Ls}
    L_{s} = 4\pi R_{2}^{2} \sigma T_{s}^{4},
\end{equation}
where $\sigma$ is the Stefan-Boltzmann constant and $T_{s}$ is the surface temperature. In order to determine the surface temperature, we must employ some boundary condition. We found the simplest way to create such a boundary condition was to directly compute surface temperatures of mass losing stars from detailed stellar evolution codes. Using the evolution code DEuCES \citep[DElaware Code for Evolution of Stars, see][and references therein for details]{DEuCES} we evolved stars from a starting mass of 0.4 $M_{\odot}$ losing mass at a constant rate. \revII{The outer boundary conditions used in the DEuCES models are the standard photospheric boundary conditions (see Chapter 5.2 in \cite{Jim_book}). For the opacities we use a blend of \cite{Ferguson_opacity}, \cite{Freedman_1} and \cite{Freedman_opacity_2}. We adopt the mixing length theory of \cite{mihalas_MLT}.} The DEuCES models have a hydrogen mass fraction of $X = 0.71310$, a helium mass fraction of $Y = 0.27029$, a metal mass fraction of $Z = 0.01661$ and a mixing length alpha of 2. These models were conducted for four constant mass loss rates: $\dot{M}_{2} = 10^{-12}$, $10^{-11}$, $10^{-10}$ and $10^{-9}$ $M_{\odot}$/yr. At the lowest mass loss rate of $\dot{M}_{2} = 10^{-12}$ $M_{\odot}$/yr, composition changes due to nuclear reactions were disabled as the nuclear timescale was short enough compared to the mass loss timescale to be relevant. From this we have the surface temperature computed on a grid of $\log \dot{M}_{2}$ and $M_{2}$ values. We then linearly interpolated over this grid to obtain the surface temperature as a function of mass-loss rate and mass of the secondary. 

This procedure was also carried out \revII{in DEuCES} with models including the magnetic inhibition of convection. The composition and mixing length alpha are the same as the previous case and the magneto-convection parameters are $\delta = 0.2$ and $B_{\text{ceil}} = 10^{4}$ G. This choice of ceiling value is motivated by arguments that the maximum attainable field strength in the interior of low mass stars is around $10-20$ kG \citep[see][and references therein]{MM17}. With this, we have two boundary conditions, one with the effects of the magnetic inhibition of convection and one without.

\subsection{Evolution of the Binary System and Mass Loss from the Secondary}

The angular momentum of a binary system in a circular orbit is:

\begin{equation}\label{Eq: J}
    J = M_{1}M_{2}\sqrt{\frac{Ga}{M}},
\end{equation}
where $M_{1}$ is the mass of the WD primary, $M = M_{1} + M_{2}$ is the total mass of the binary and $a$ is the binary separation. The binary separation and the orbital period ($P$) are related through Kepler's third law:

\begin{equation}
    P^{2} = \frac{4\pi^{2}}{GM}a^{3}.
\end{equation}
The AML comes in two forms: systemic AML, such as gravitational radiation and magnetic braking, and CAML, which is associated with the mass transfer process:

\begin{equation}
    \frac{\dot{J}}{J} = \frac{\dot{J}_{\text{sys}}}{J} + \frac{\dot{J}_{\text{CAML}}}{J}.
\end{equation}
CAML is typically parameterized in the following way \citep{CAML_param}:

\begin{equation}\label{eq: J dot CAML}
    \frac{\dot{J}_{\text{CAML}}}{J} = \nu \frac{\dot{M}_{2}}{M_{2}},
\end{equation}
where $\nu>0$ is a parameter measuring the amount of AML due to the mass transfer process. Mass transfer occurs when the radius of the secondary is larger than the effective radius of the Roche lobe of the secondary. For the Roche lobe radius, we use Eggleton's relation \citep{RL_egg}:

\begin{equation}
    R_{L} = \frac{0.49 q^{2/3}}{0.6 q^{2/3} + \ln\left(1 + q^{1/3}\right)} a.
\end{equation}
Once mass transfer begins, the radius of the secondary is forced to evolve along with the Roche lobe radius: $R_{2} = R_{L}$. We define $q = M_{2} / M_{1}$ as the mass ratio of the binary and $\beta = \dot{M} / \dot{M}_{2}$ as the ratio of the total mass loss rate of the system to the mass loss rate of the secondary. Combining all the above equations, we arrive at an expression for the mass loss rate of the secondary\footnote{The expression for $D$ displayed here is different from the one appearing in \cite{CAML_param}. This expression was derived using the Eggleton relation for the Roche lobe radius while \cite{CAML_param} used the \revI{simpler} Paczynski relation. Additionally, \cite{CAML_param} include the mass-radius exponent in their expression for $D$. In our expression, reference to the change in radius has been pulled out, therefore $D$ is a function only of the binary variables $q$, $\beta$ and $\nu$.}:

\begin{equation}\label{eq: M dot 2}
    \frac{\dot{M}_{2}}{M_{2}} = \frac{1}{D}\left[\frac{\dot{J}_{\text{sys}}}{J} - \frac{\dot{R}_{2}}{2 R_{2}}\right],
\end{equation}
where

\begin{equation}
    D = \frac{1 + 0.5\beta q + \left(\beta - 1\right)q^{2}}{1 + q} - \frac{C}{2}\left[q + \left(1-\beta\right)q^{2}\right] - \nu,
\end{equation}

\begin{equation}
    C = \frac{2}{3q} - \frac{1.2 + 1.2q^{-1/3} + q^{-2/3}}{3\left(1 + q^{1/3}\right)\left[0.6q^{2/3} + \ln\left(1+q^{1/3}\right)\right]}.
\end{equation}
Equation \ref{eq: M dot 2} can be employed \enquote{as is} with $\dot{R}_{2}$ being computed numerically over each timestep. For the polytrope model, equation \ref{Eq: mass rad relation} relates the radius to the mass. Thus, for the $n=3/2$ polytrope specifically, the mass loss rate of the secondary is related directly to the change in polytropic constant:

\begin{equation}\label{Eq: M dot poly}
    \frac{\dot{M}_{2}}{M_{2}} = \left(D - \frac{1}{6}\right)^{-1} \left(\frac{\dot{J}_{\text{sys}}}{J} - \frac{\dot{K}}{2K}\right).
\end{equation}
Lastly, we need an expression for the systemic angular momentum loss. For CVs \revII{below the gap}, we consider only gravitational radiation \revII{(in Section \ref{sec: additional AML} we will consider some additional AML mechanisms)}. The AML rate due to gravitational radiation is given by \revI{\citep[see][]{J_dot_GR}}:

\begin{equation}\label{eq: J dot GR}
    \dot{J}_{\text{GR}} = -\frac{32}{5} \frac{G^{7/2}}{c^{5}}\frac{M_{1}^{2}M_{2}^{2}M^{1/2}}{a^{7/2}},
\end{equation}
where $c$ is the speed of light.

\subsection{Numerical Procedure}

All of our models start with a secondary mass of 0.3 $M_{\odot}$ at an orbital period of 4 hours. The composition is the same as the detailed models discussed previously: $X = 0.71310$, $Y = 0.27029$ and $Z = 0.01661$. Unless otherwise stated, the mass of the WD is $0.81$ $M_{\odot}$ which is the average observed WD mass in CVs \citep[see][]{ave_WD_mass}. For the binary parameters, we assume that novae outbursts carry away all the mass transferred from the secondary, implying $\beta = 1$. \revII{For our first set of models,} we do not include CAML, thus $\nu = 0$. We did compute models with the standard form of CAML which states that novae outbursts carry away the specific angular momentum of the white dwarf, implying that $\nu = M_{2}^{2}/\left(M_{1}\left(M_{1}+M_{2}\right)\right)$ \citep{CAML_param}.  However we found that the inclusion of CAML in this form had negligible impacts around the period minimum, and only minor impacts for larger secondary masses. The AML rate for CAML was always an order of magnitude smaller than the AML rate due to gravitational radiation, and over two orders of magnitude smaller around the period minimum (a result also found by \cite{Knigge}, their Tables 3 and 4). Therefore, whether CAML in the standard form is included or not does not impact the results of this study. \revII{In Section \ref{sec: additional AML}, we will examine the effects of the empirical model for CAML (eCAML) proposed by \cite{CAML}.}
 
With these starting conditions, the Roche lobe is large enough such that mass transfer does not occur immediately. An initial guess is given for the radius of the secondary which is converted to an initial value for $K$ with equation \ref{Eq: mass rad relation}. Since mass transfer has not yet begun, the secondary is allowed to relax into thermal equilibrium. To compute the surface temperature for the initial state with no mass loss, we use the value of the interpolation at the starting mass and lowest mass loss rate. The binary separation will shrink due to gravitational radiation until $R_{L} < R_{2}$, where mass transfer will begin. For \revI{determining} the timestep, we consider the thermal and mass-loss timescales. For an $n=3/2$ polytrope, the thermal timescale is:

\begin{equation}
    \tau_{\text{th}} = \frac{6}{7} \frac{G M_{2}^{2}}{R L_{s}},
\end{equation}
and the mass-loss timescale is:

\begin{equation}
    \tau_{\dot{M}} = \frac{M_{2}}{\dot{M}_{2}}.
\end{equation}
The timestep was required to be 1\% \revI{of} the smaller of the two timescales: $dt = 0.01 \min\left(\tau_{\text{th}}, \tau_{\dot{M}}\right)$. The simulation was run until the mass of the secondary fell below 0.03 $M_{\odot}$. Computed values are output after every $10^{8}$ years or after every $0.001$ $M_{\odot}$ lost by the secondary. 

\section{Results} \label{sec:Results}

\subsection{Impact of Magneto-Convection on System Properties}

\begin{figure}
    \centering
    \includegraphics[width=\linewidth]{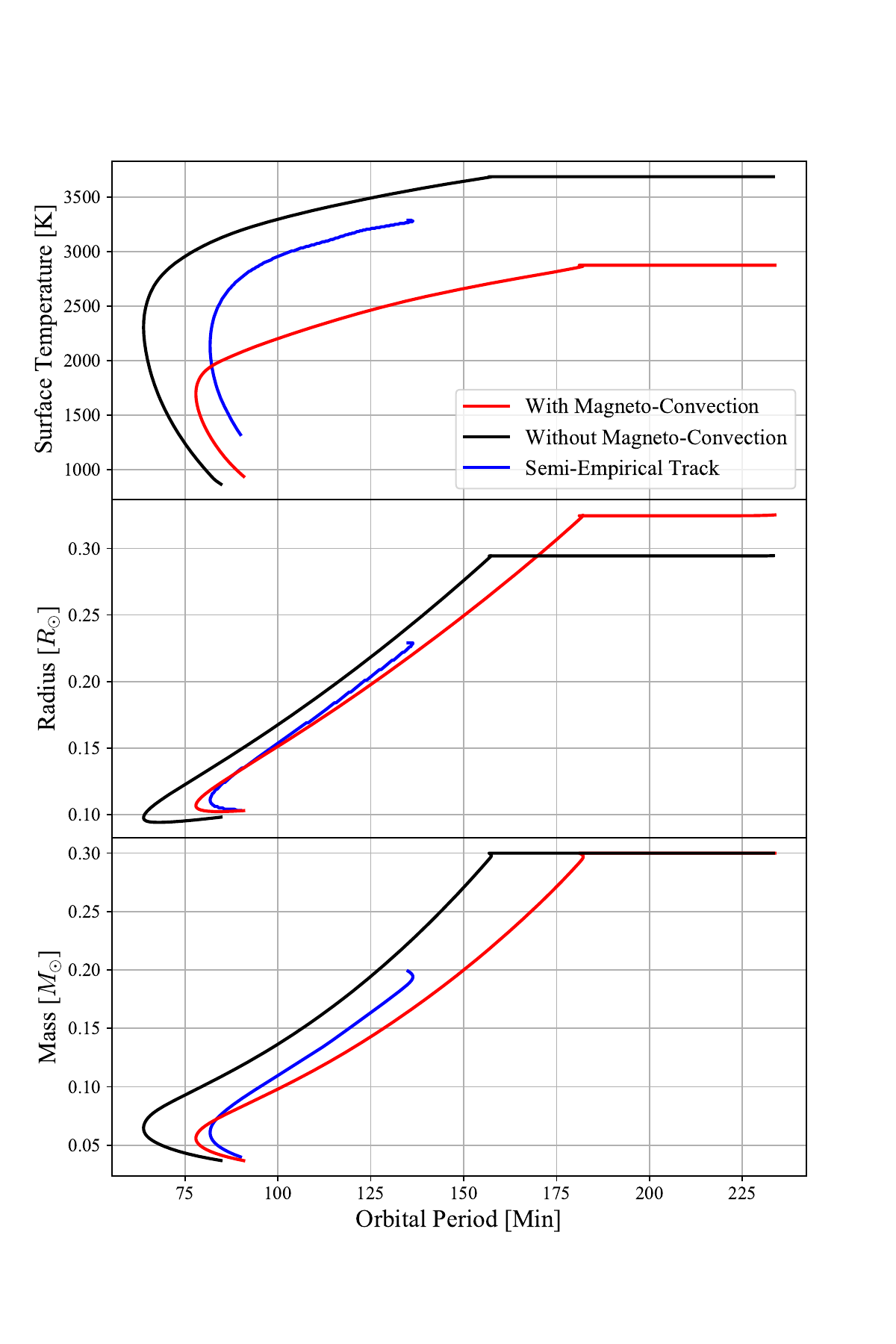}
    \caption{Properties of the model system, including secondary surface temperature (top), secondary radius (middle) and \revII{secondary mass} (bottom) vs. \revII{orbital period}. The black curves displays the results using the non-magnetic boundary condition while the red curve displays the results using the magnetic boundary condition. \revII{The blue curve is the updated semi-empirical donor sequence of \cite{Knigge}.}}
    \label{fig: Secondary Properties}
\end{figure}

Figure \ref{fig: Secondary Properties} displays properties of the model system vs. \revII{the orbital period}. Note that since the properties are plotted against the \revII{orbital period}, the evolution in time \revII{follows the curves starting from the right (at large orbital period)}. For both the magnetic and non-magnetic models, mass transfer decreases the secondary mass leading to a decrease in surface temperature. The secondary radius also decreases, except for at the smallest secondary masses, where the radius levels off and begins to increase due to the secondary being driven out of thermal equilibrium and growing electron degeneracy. The period also decreases for most of the evolution, but the period bounce stage is clearly evident.

We find that the magnetic models are cooler than the non-magnetic models except for the smallest masses. Recall that these temperatures are computed directly from the employed boundary condition. Due to the cooler temperatures, the magnetic models are larger. Thus the inclusion of magneto-convection increases the radii of the secondary (an expected result based on the known effects of magneto-convection). A consequence of the larger secondary is that the binary separation, and thus the orbital period, are larger for the magnetic models. This leads to an increase in the minimum period by $14$ minutes, from 63.9 minutes for the non-magnetic models to 77.9 minutes for the magnetic models.

\begin{figure}
    \centering
    \includegraphics[width=\linewidth]{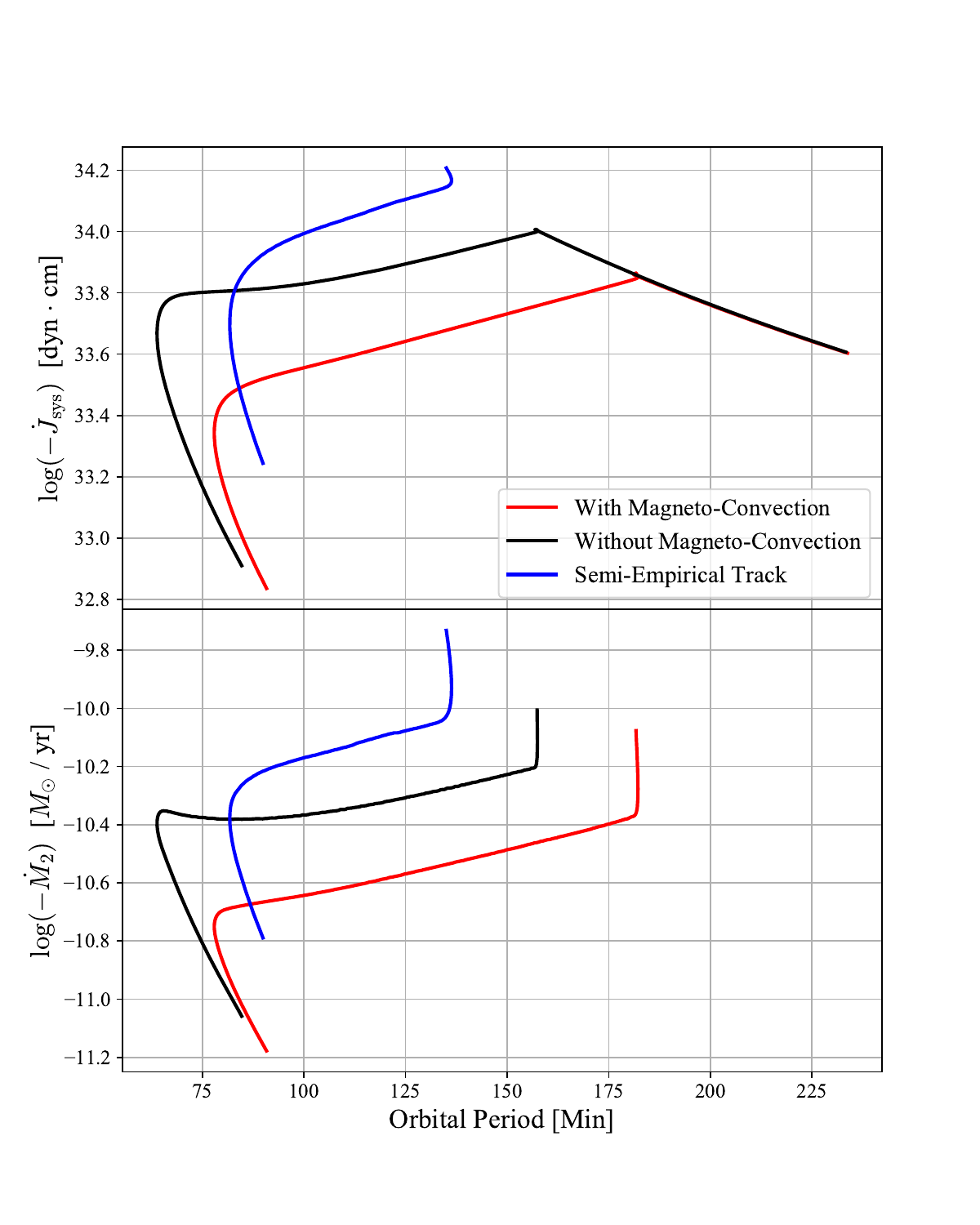}
    \caption{Angular momentum loss rate (top) and mass-loss rate (bottom) vs. \revII{orbital period}. The black curves displays the results using the non-magnetic boundary condition while the red curve displays the results using the magnetic boundary condition. \revII{The blue curve is the updated semi-empirical donor sequence of \cite{Knigge}.}}
    \label{fig: rates}
\end{figure}

Figure \ref{fig: rates} displays the AML rate and mass-loss rate vs. \revII{orbital period}. Prior to the onset of mass loss, the AML rate increases due to the decreasing binary separation. Recall that the AML is solely due to gravitational radiation and is described by equation \ref{eq: J dot GR}. Once mass loss begins, the mass-loss rate is strong enough to overpower the decreasing binary separation, leading to a decrease in the AML rate. However, due to the competition between the decreasing binary separation and decreasing mass, the AML rate does not substantially decrease between the onset of mass transfer and the period minimum. \revII{This interplay between the decreasing binary separation and decreasing secondary mass leads to a slight leveling off in the non-magnetic case just prior to the period minimum.} \revII{At the period minimum and during the period bounce phase}, the AML rate drops rapidly as the decreasing mass and increasing binary separation both lead to a decreasing AML rate. The evolution of the mass-loss rate mimics the evolution of the AML rate. The small bump in the mass loss rate \revII{just before the} period minimum for the non-magnetic case is caused by the leveling off of the AML rate. While $\dot{J}_{\text{sys}}$ remains constant, $J$ still decreases, therefore $\dot{J}_{\text{sys}}/J$ increases leading to the small bump in the mass loss rate.

Comparing the magnetic and non-magnetic models, we find that the magnetic models always have a smaller AML rate. Adding magneto-convection increases the orbital period, and thus the binary separation. Increasing the binary separation decreases the AML rate due to gravitational radiation. Therefore, throughout the evolution, gravitational radiation is weaker in the magnetic models leading to a decreased AML rate. This decreased AML rate also leads to smaller mass loss rates for the magnetic models.

\revII{Included in both Figures \ref{fig: Secondary Properties} and \ref{fig: rates} are the updated semi-empirical evolution tracks of \cite{Knigge} for below the period gap. The sequences produced by \cite{Knigge} are full evolution tracks, starting above the period gap. The high mass transfer rates above the period gap, driven by magnetic braking, delay the transition to fully convective until 0.2 $M_{\odot}$, explaining why these tracks begin at lower masses then our models. Regardless, we find that while our approximate models do not match the exact positions of the \cite{Knigge} results, they do capture the general behavior. Therefore, we can expect the conclusions drawn in this work about the effects of magneto-convection on short-period CVs to apply to other, more detailed, evolutionary models.}

\subsection{Additional Angular Momentum Loss Mechanisms}\label{sec: additional AML}

\begin{figure}
    \centering
    \includegraphics[width=\linewidth]{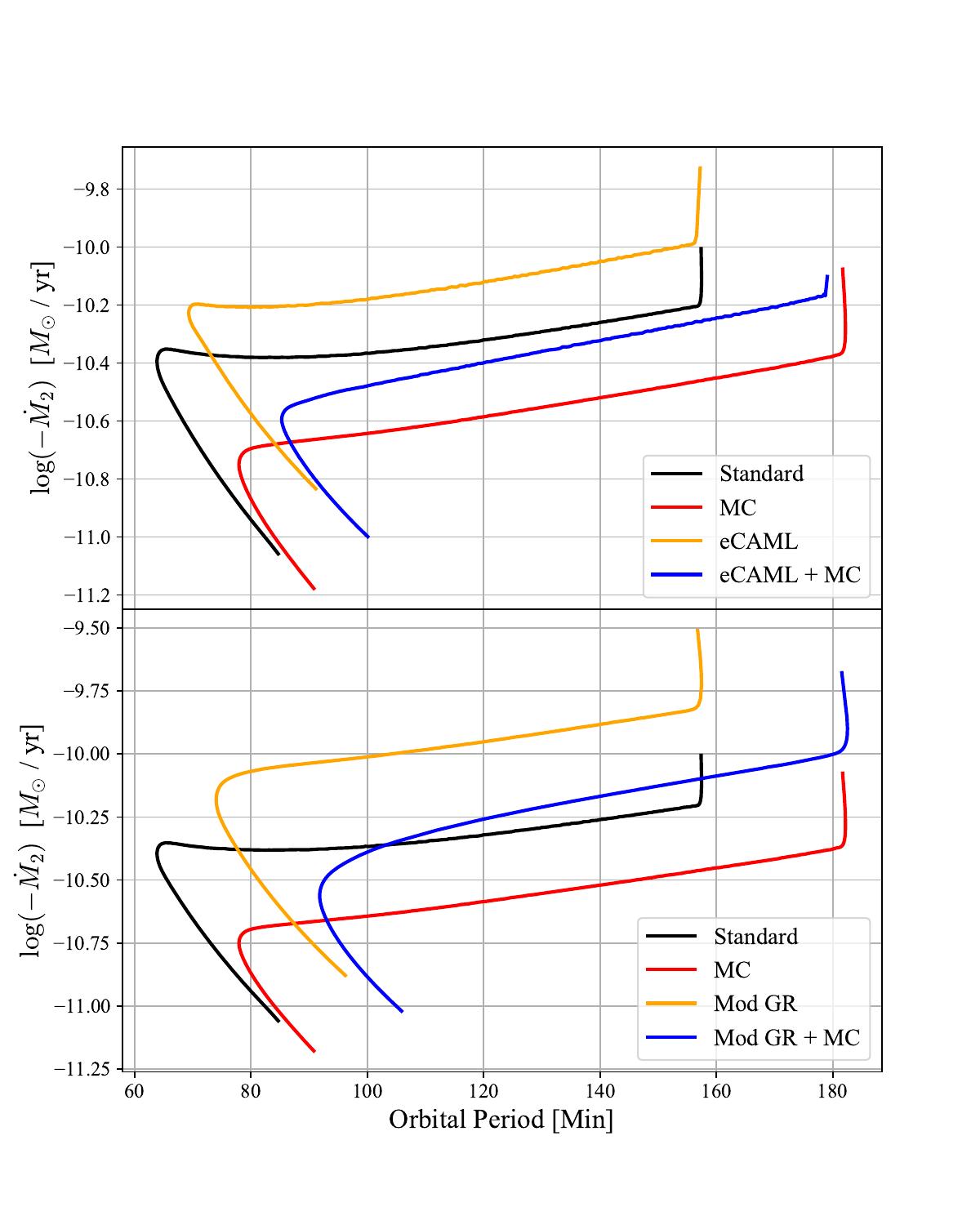}
    \caption{\revII{Mass-loss rate vs. orbital period for the cases with additional AML mechanisms. The top plot shows the results with the eCAML model and the bottom plot shows the results with modified gravitational radiation (Mod GR). In both plots we also include the standard model with only gravitational radiation and the standard model with magneto-convection (MC).}}
    \label{fig: additional AML}
\end{figure}

\revII{To examine the effects of magneto-convection at larger AML rates, we conducted simulations with additional AML mechanisms. First is the eCAML model proposed by \cite{CAML}, which assumes the following functional form for the $\nu$ parameter appearing in equation \ref{eq: J dot CAML}}:

\begin{equation}
    \nu = \frac{C}{M_{1}}.
\end{equation}
\revII{\cite{CAML} find that this functional form with $C = 0.3 - 0.4$ $M_{\odot}$ produces good agreement between observations and simulated CV populations. For this study we adopt $C = 0.35$ $M_{\odot}$. We also modified the AML rate due to gravitational radiation. \cite{Knigge} have shown that better agreement with observations below the period gap is found when the AML rate due to gravitational radiation is scaled by a factor of $\sim2.5$. The physical mechanism for this extra AML could be some residual magnetic braking which remains below the period gap.}
\revII{We re-ran the simulations now including eCAML and the scaled gravitational radiation. Figure \ref{fig: additional AML} displays plots of the mass-loss rate vs. orbital period for both cases. We re-confirm that adding either eCAML or scaled gravitational radiation increases the period minimum and the mass-loss rates when compared to a model with (standard) gravitational radiation alone. When we include magneto-convection along with the additional AML mechanism, we find that same results found in the previous section still hold. In particular, even when additional AML mechanism are present, the inclusion of magneto-convection leads to longer orbital periods (causing an increased period minimum) and lower mass-loss rates. Comparing the case of magneto-convection alone to the cases with magneto-convection plus additional AML (comparing the red and blue curves in both parts of Figure \ref{fig: additional AML}), we find that the reduced mass-loss rates caused by including magnetic effects can be compensated by including additional AML mechanisms.}

\subsection{Effect of the White Dwarf Mass}

\begin{figure}
    \centering
    \includegraphics[width=\linewidth]{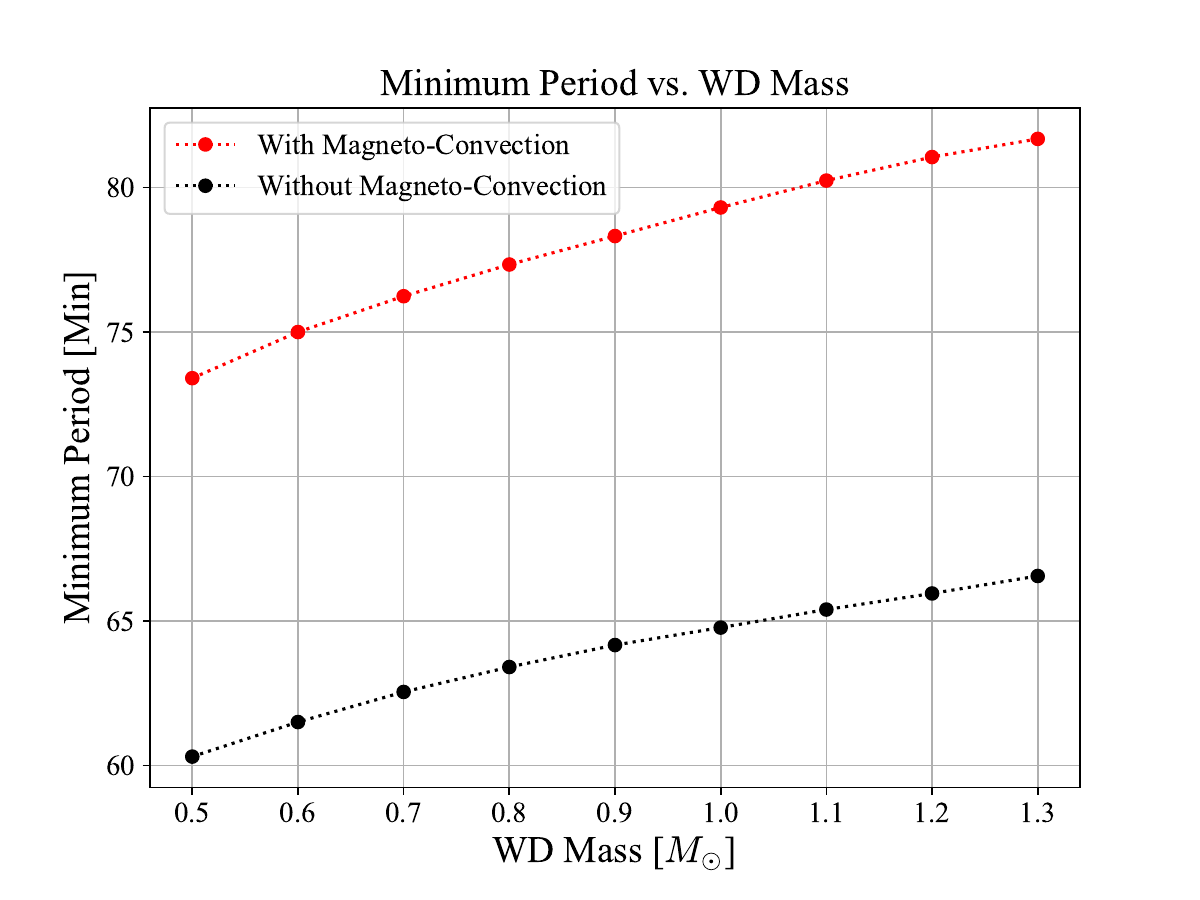}
    \caption{Minimum period reached by the simulation vs. the white dwarf mass.}
    \label{fig: min P vs WD mass}
\end{figure}

To test the impact of the white dwarf mass on the minimum period, we re-ran the simulations with white dwarf masses ranging from 0.5 - 1.3 $M_{\odot}$. Figure \ref{fig: min P vs WD mass} displays the minimum period vs the WD mass. For both the magnetic and non-magnetic models, increasing the WD mass increases the minimum period. Since $\dot{J}_{\text{GR}}$ depends on the white dwarf and total mass (see equation \ref{eq: J dot GR}), increasing the white dwarf mass increases the AML rate. An overall increase in the AML rate increases the period minimum, causing the trend in Figure \ref{fig: min P vs WD mass}. Increasing the WD mass from 0.5 - 1.3 $M_{\odot}$ increases the minimum period by 6.2 minutes for the non-magnetic models and 8.3 minutes for the magnetic models. Using a sample of 89 CVs with accurate WD mass measurements, \cite{ave_WD_mass} constrain the average WD mass to 0.61 - 0.97 $M_{\odot}$. \revII{Over the range of 0.6 - 1.0 $M_{\odot}$, the change in the period minimum is only 3.3 minutes for the non-magnetic models and 4.3 minutes for the magnetic models.} Based on the minimal changes in the period minimum over this range of WD mass, increasing the WD mass is unlikely to explain the period minimum problem.

\section{Discussion} \label{sec:discussion}

The assumed strong magnetic braking in long period CVs requires the MS secondaries to be magnetic. Based on mounting evidence that fully convective stars can maintain strong magnetic fields, magnetic effects may be relevant for short period CVs as well. The inclusion of magneto-convection has major impacts on short period CV evolution. The inhibition of convection due to magnetic fields causes the MS secondaries to be inflated in size and cooler in temperature. The larger secondaries lead to larger binary separations, larger orbital periods and smaller mass-loss rates. These effects raise the minimum period by 14 minutes. Thus, magneto-convection can help alleviate the period minimum problem. 

On the other hand, an increase in the WD mass is unlikely to alleviate the period minimum problem. Over the range for which \cite{ave_WD_mass} constrain the average WD mass, the period minimum only changes by $\sim3$ minutes for the non-magnetic models. \cite{ave_WD_mass} also examine the WD mass distribution, finding no evidence for a dependence of the WD mass on the orbital period. This observational evidence hints that the WD does not increase in mass over the evolution of the CV. Nevertheless, WD mass growth in CVs remains a debated topic, particularly in the context of type 1a supernovae. Based on the small change in the period minimum over the tested WD masses, it is unlikely that a growing WD mass could solve the period minimum problem (if the WD grows in mass at all). 

It should be stated that our models are limited by the $n=3/2$ polytrope approximation used for the fully convective, MS secondaries. The goal of this study was to compare the magnetic and non-magnetic models to understand the effect of magneto-convection. From this comparative study, we find that magneto-convection can substantially increase the period minimum \revII{while also decreasing the mass-loss rate}. \revII{This is a first step to fully understanding the effects of magneto-convection in CVs. In the future, the use of detailed stellar evolution codes will help reach further conclusions about the role of these magnetic effects. For instance, in this study we adopted one particular combination of values for $\delta$ and $B_{\text{ceil}}$. Detailed codes could explore the full $\delta-B_{\text{ceil}}$ parameter space to better understand the impact of magneto-convection on the entire population of CVs.} 

A detractor for including magneto-convection is the decrease in the AML and mass transfer rate. Due to compressional heating of the WD from the accretion disk, WD effective temperatures provide a measurement of the mass transfer rate \citep{Teff_to_Mdot}. Using this, \cite{pala_Teff} have shown that CVs below the period gap have larger WD effective temperatures, and thus larger AML and mass transfer rates, then predicted by standard evolution tracks. Similar results have also been reported by \cite{ave_WD_mass} through direct measurements of mass transfer rates (rather than indirect inferences through WD effective temperatures). While the magnetic effects lead to a smaller AML and mass-loss rate, the inclusion of magneto-convection does not preclude the addition of other AML loss mechanisms. In particular, if the magnetic field remains, some form of magnetic braking may persist under the period gap. \revII{As demonstrated in Figure \ref{fig: additional AML}, additional AML mechanism can help compensate for the reduced mass-loss rates caused by magneto-convection.}

The main result of this study demonstrates that magneto-convection does have an effect on short period CVs. If the low mass, fully convective secondaries are presumed to be magnetic, then these effects should be considered in models. In addition, magneto-convection should also be relevant for long period CVs where magnetic braking is already presumed to be the dominant AML mechanism. Models presented by \cite{MM24} have shown that the relative radius inflation of fully convective stars is smaller than for stars with a radiative core \citep[see also][]{other_rad_infl}. The decrease in relative radius inflation at the fully convective transition could be important for the period gap.

\section{Conclusion} \label{sec:conclusion}

We have built a simple model to demonstrate the effects of magneto-convection on short period CVs. Our simulations show that the inclusion of magneto-convection causes inflated secondaries, leading to an increase in the orbital period and a decrease in the mass-loss rate. \revII{We also verify that these conclusions hold when additional AML mechanisms act.} The minimum orbital period is increased by 14 minutes, indicating that the inclusion of magneto-convection can help alleviate the period minimum problem. In addition, we find that the WD mass does not substantially increase the minimum period, thus it will be unlikely to explain the period minimum problem with a larger average WD mass or with WD mass growth. These results demonstrate that magneto-convection does have an impact on CV evolution. If the MS secondaries are presumed to be magnetic, then the effects of these magnetic fields on stellar structure should be incorporated into models.\\

C.L. graciously acknowledges support from the Delaware Space Grant College and Fellowship Program (NASA Grant 80NSSC20M0045). C.L. would like to thank Edward Sion and J.M. would like to thank Dermott Mullan for their informative insights and discussions.

\bibliography{Bibliography.bib}{}
\bibliographystyle{aasjournal}

\end{document}